\documentclass[12pt]{spieman}  
\usepackage{amsmath,amsfonts,amssymb}
\usepackage[pdftex]{graphicx}
\usepackage{setspace}
\usepackage{tocloft}

\title{Pile-up simulator for XRISM/Xtend onboard the X-ray Imaging and Spectroscopy Mission (XRISM)}

\author[a,*]{Tomokage Yoneyama}
\author[a]{Tsubasa Tamba}
\author[b]{Hirokazu Odaka}
\author[c]{Aya Bamba}
\author[d]{Hiroshi Murakami}
\author[e]{Koji Mori}
\author[f]{Yukikatsu Terada}
\author[g]{Masayoshi Nobukawa}
\author[h]{Tsunefumi Mizuno}
\affil[a]{Japan Aerospace Exploration Agency, Institute of Space and Astronautical Science, 3-1-1 Yoshino-dai, Chuo-ku, Sagamihara, Kanagawa 252-5210, Japan}
\affil[b]{Osaka University, Department of Earth and Space Science, 1-1 Machikaneyama-cho, Toyonaka, Osaka 560-0043, Japan}
\affil[c]{University of Tokyo, School of Science, 7-3-1 Hongo, Bunkyo, Tokyo 113-0033, Japan}
\affil[d]{Tohoku Gakuin University, Faculty of Informatics, 3-1 Shimizukoji, Wakabayashi-ku, Sendai, Miyagi 984-8588, Japan}
\affil[e]{University of Miyazaki, Faculty of Engineering, 1-1 Gakuen Kibanadai Nishi, Miyazaki 889-2192, Japan}
\affil[f]{Saitama University, Graduate School of Science and Engineering, 255 Shimo-Okubo, Sakura, Saitama, Saitama 338-8570, Japan}
\affil[g]{Nara University of Education, Faculty of Education, Takabatake-cho, Nara, Nara 630-8528, Japan}
\affil[h]{Hiroshima University, School of Science, 1-3-1 Kagamiyama, Higashi-Hiroshima, Hiroshima 739-8526, Japan}

\cftpagenumbersoff{figure}
\cftpagenumbersoff{table} 
\begin{document} 
\maketitle

\begin{abstract}
In X-ray astronomy, most observatories utilize multi-pixel photon-counting devices. When a photon-counting device observes a bright source, we face an unavoidable problem called pile-up. Pile-up leads to mistakes in the observational properties of the source, mainly an apparent decrease in the X-ray flux. X-Ray Imaging and Spectroscopy Mission ({\it XRISM}) has two X-ray telescopes, one of which is Xtend, a CCD camera with a wide field-of-view (FOV) of 38 arcmin square. Xtend has three operating modes: full window mode with a frame exposure of $\sim$4 sec, 1/8 window mode with $\sim$0.5\,sec and reduced FOV, and 1/8 window mode with burst option, whose frame exposure is reduced to $\sim$0.06 s. Observers need to select the operating mode according to their target fluxes. We develop the pile-up simulator for Xtend to provide a quantitative assessment of pile-up according to the fluxes, spectra, and shapes of X-ray sources. We derived the 10\% pile-up limits for a point source of 7.8, 66.2, and 447.9\,counts\,s$^{-1}$ for full window, 1/8 window, and 1/8 window mode with burst option, respectively, by assuming the Crab spectrum. We present further simulations for a diffuse source and monochromatic spectra.
\end{abstract}


{\noindent \footnotesize\textbf{*}Tomokage Yoneyama,  \linkable{yoneyama.tomokage@jaxa.ac.jp} }

\begin{spacing}{2}   

\section{Introduction}
\label{sec:intro}  

Most recent X-ray astronomical satellites equip photon-counting devices as focal plane detectors of the X-ray telescopes to perform imaging spectroscopy and timing analysis. Once an X-ray photon enters a photon-counting device, an electron cloud is generated by the photoelectric effect. Because of the large energy of the X-ray photon, the electron cloud may spread over several pixels, from which an X-ray event is reconstructed. The signal patterns of X-ray photons on the frame image must be distinguished from those of non-X-ray signals, such as cosmic rays. The so-called Grade algorithm introduced for {\it ASCA}/SIS\cite{burke91} has been employed to classify whether the extracted events are caused by incident X-ray photons or not. When the Grade algorithm is applied, a photon-counting device assumes that only one photon is detected in an adjacent 3$\times$3 (or 5$\times$5) pixels within a single readout frame. If the device observes an X-ray source bright enough to violate this assumption, more than two photons may enter the same pixel or several nearby pixels within a single readout. This is called pile-up, an unavoidable issue for photon-counting devices. Once a pile-up occurs, multiple photons are detected as a single photon event with their combined energy, or as a bad-grade event. This leads to mistakes in the observational properties of the source, mainly an apparent decrease in the X-ray flux and distortion of the spectral shape, which mainly appears as a hardening. X-ray telescopes utilizing CCDs as multi-pixel photon-counting devices for imaging and spectroscopy, such as {\it ASCA}\cite{tanaka94},  {\it Chandra}\cite{weisskopf02}, {\it XMM-Newton}\cite{jansen01}, and {\it Suzaku}\cite{mitsuda07}, have operating modes to mitigate the pile-up for bright sources.

Many studies have been presented to attribute the pile-up. Ballet (1999)\cite{ballet99} and Ballet (2003)\cite{ballet03} reported theoretical studies for pile-up focusing on detectors with small pixels compared to the point spread function (PSF) of X-ray mirrors. They took only single-pixel events into account. Davis (2001)\cite{davis01} presented an analysis model for X-ray spectra affected by the pile-up, introducing grade-migration probability. The model does not consider the energy dependence of the pile-up, which is in reality significant. Recently, Sevilla (2017)\cite{sevilla17} constructed an analytic model of the pile-up considering the probability that a detector observes $n$ counts if $N$ particles are incident in a single exposure frame when incident photons have a Poisson distribution. They applied the model to the XMM-Newton observation of an X-ray pulsar. Dauser et al. (2019)\cite{dauser19} presented a mission-independent simulation toolkit, the simulation of X-ray telescopes (SIXTE), which was applied to study the pile-up of Athena-WFI.

X-Ray Imaging and Spectroscopy Mission ({\it XRISM})\cite{tashiro24} launched on 2024-09-07 equips two X-ray telescopes, Resolve\cite{ishisaki18} and Xtend\cite{mori24}. Xtend is a combination of a Wolter I-type X-ray telescope, X-ray Mirror Assembly (XMA), and a CCD camera array, Soft X-ray Imager (SXI), whose specification is summarized in table \ref{tab:spec}. Xtend performs a wide-field X-ray imaging spectroscopy with a field of view (FOV) of 38.5\,arcmin square and a wide bandpass of 0.4 -- 13.0\,keV with energy resolution of $\sim$ 180\,eV at 6\,keV. Three operating modes are implemented for Xtend: full window mode with a frame exposure of 4 sec, 1/8 window mode with 0.5 sec and reduced FOV, and 1/8 window mode with a 0.1 s burst option. Specification of each mode is summarized in table \ref{tab:mode}. Observers should select an appropriate mode that meets their scientific purposes and the source brightness to minimize the pile-up. We develop the pile-up simulator for Xtend to provide a quantitative assessment of pile-up according to the fluxes, shapes, and spectra of X-ray sources.  In this paper, we report the development of the pile-up simulator, its application, and the anticipated effects of pile-up for Xtend.

\begin{table}[ht]
\caption{Specification of Xtend/SXI CCDs} 
\label{tab:spec}
\begin{center}       
\begin{tabular}{ll} 
\hline
CCD Type & Frame transfer, p-channel, Backside illuminated  \\
Number of CCDs & 4 \\
Number of segments per CCD & 2 \\
Imaging area & 31 $\times$ 31\,mm$^2$, 640 $\times$ 640 logical pixels \\
Pixel size & 48 $\times$ 48\,$\mu$m (logical pixel) \\
Plate scale & 1$^{\prime\prime}$.77\,pix$^{-1}$ (logical pixel) \\
Depletion depth & 200\,$\mu$m \\
Optical Blocking Layer (OBL) & Al (200\,nm) on the CCD surface \\
Contamination Blocking Filter (CBF) & Al (80\,nm)/Polyimide (200\,nm)/Al (40\,nm) \\
Nominal operating temperature & $-$110 $^\circ$C \\
\hline
\end{tabular}
\end{center}
\end{table}

\begin{table}[ht]
\caption{Available operation modes of Xtend} 
\label{tab:mode}
\begin{center}       
\begin{tabular}{cccccc} 
\hline
 & Image Size$^*$ & Time Res. & Frame Exposure & LTF$^\dagger$  \\
 & (pixels $\times$ pixels) & (sec) & (sec) & \\
 \hline
Full Window & 640 $\times$ 640 & 4 & 3.9631 & 0.99 \\
1/8 Window & 640 $\times$ 80 & 0.5 & 0.4631 & 0.93 \\
1/8 Window with 0.1\,s Burst & 640 $\times$ 80 & 0.5$^\ddag$ & 0.0620 & 0.12 \\
\hline
\end{tabular}
\item[$^\dagger$] Live time fraction
\item[$^\ddag$] Photon arrival times are determined with 0.062 sec accuracy.
\end{center}
\end{table}

\section{PILE-UP SIMULATOR}
\label{sec:pusim}

The Xtend pile-up simulator is based on Tamba et al. (2022)\cite{tamba22}, which is the simulation framework for {\it Suzaku}/XIS. It simulates the physical processes of how incident X-ray photons interact in a detector, appear as signals on a frame image, and are extracted as events. All functions constructing the simulator are implemented in a framework for X-ray detector and data analysis simulation, ComptonSoft (Odaka et al. 2010\cite{odaka10}). ComptonSoft performs Monte Carlo simulations of X-ray photons interacting with detectors based on Geant4 (Agostinelli et al. 2003\cite{agostinelli10}; Allison et al. 2016\cite{allison16}), a library of high-energy particle physics simulation. In this paper, we employ Comptonsoft 5.7.0 and Geant4 10.05.01.

We construct a physical model of Xtend/SXI based on its specifications, as shown in table \ref{tab:spec}. Simulation parameters are determined to reproduce the on-ground calibration of Xtend/SXI reported in Yoneyama et al.(2021).\cite{yoneyama21}\cite{mori22}. A significant specification of a detector to determine the pile-up tolerance is the energy-dependent grade distribution, which is determined by the charge cloud diffusion during its drift toward the electrodes. The diffusion is controlled by the vertical structure of the electric field in the depletion layer; the stronger field achieves a fast drift, resulting in less diffusion. Hence, what is important to simulate pile up is to construct the electric field structure that reproduces the energy-dependent grade distribution. In this simulator, the electric field structure is empirically parametrized as a linear function along the depth of the depletion layer (see 3.2.2 in Tamba et al. 2022\cite{tamba22}). We tuned the parameters to reproduce the grade distribution of Xtend/SXI for the emission lines: O K$\alpha$, F K$\alpha$, Al K$\alpha$, Si K$\alpha$, Mn K$\alpha$, and K$\beta$, respectively obtained in the ground calibration. The event and split threshold are 198 and 90\,eV, respectively, which are the nominal ones for Xtend.
The simulator utilizes the {\it XRISM} calibration database (CALDB) to simulate the properties of Xtend, e.g., PSF. In this simulation, charge transfer after a frame exposure and charge transfer inefficiency (CTI) are not reproduced because they only affect the spectroscopic performance, and their influence on photon counting is considered to be sufficiently small. Accordingly, simulation of the pile-up is independent of CTI. Xtnd/SXI utilizes the charge injection method to suppress CTI. The simulator does not take the charge injection into account, because it also affects only the spectroscopic performance. The simulator takes two items as inputs: an X-ray spectrum model and an image of the X-ray source. The outputs are simulated X-ray events and spectra extracted from the events. Comparing the input and output spectra, we calculate the pile-up fraction $F_{\rm p}$ defined as

\begin{equation}
F_{\rm p} = 1 - r_{\rm out} / r_{\rm in},
\end{equation}

\noindent where $r_{\rm in}$ and $r_{\rm out}$ are input and output count rates in an arbitrary energy range, respectively. 

\section{SIMULATIONS}
\label{sec:simulations}

The Xtend pile-up simulator is applied to derive the pile-up fraction about the incident X-ray flux for several setups: a point source with the Crab spectrum, monochromatic X-rays of 1\,keV and 6\,keV, and a diffuse source with the Crab spectrum. Note that the pile-up fraction is calculated from good-grade spectra (Grade = 0, 2, 3, 4, 6, and 8; for the definition of grades, see Tamba et al. 2022\cite{tamba22}). We employed the IACHEC model (Weisskopf et al. 2010\cite{weisskopf10}) for the Crab spectrum. It is described as ``TBabs*Powerlaw'' in XSPEC, whose parameters are the hydrogen column density $N_{\rm H}$ of $3.8 \times 10^{21}$\,cm$^{-2}$, the photon index $\Gamma$ of $2.1$, and the normarization of $8.9$\,counts\,s$^{-1}$ at 1\,keV. In the TBabs model, we employed ``angr'' (Anders and Grevesse 1989\cite{anders89}) and ``vern'' (Verner and Yakovlev 1995\cite{verner95}, and Verner et al. 1996\cite{verner96}) models for abundances and cross sections, respectively.

\subsection{Point source}

\subsubsection{Pile-up fraction}
\begin{figure} [htbp]
   \begin{center}
   \includegraphics[width=10cm]{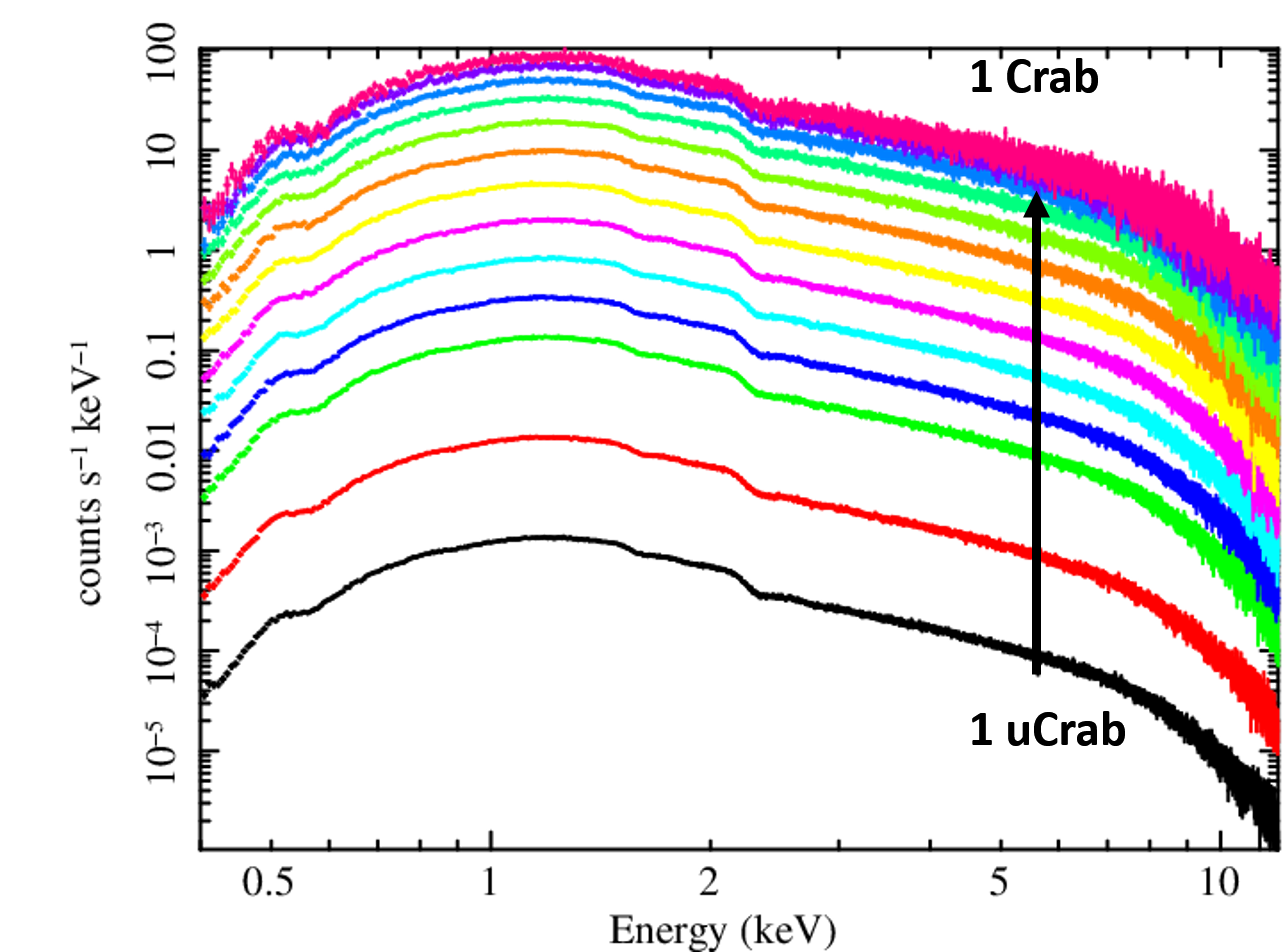}
   \end{center}
   \caption[Simulated spectra of the Crab spectrum with different fluxes] 
   { \label{fig:point_crab_spec} 
Simulated spectra by the pile-up simulator by assuming a point source with the Crab spectrum with fluxes of $10^{-6}$--$1$\,Crab. The full window mode is simulated in this figure. High input fluxes result in the high pile-up effect, by which fractions of photons are lost in the simulated spectra (see figure \ref{fig:point_crab} for the quantitative evaluation of the pile-up).}
\end{figure}

\begin{figure} [htbp]
   \begin{center}
   \includegraphics[width=10cm]{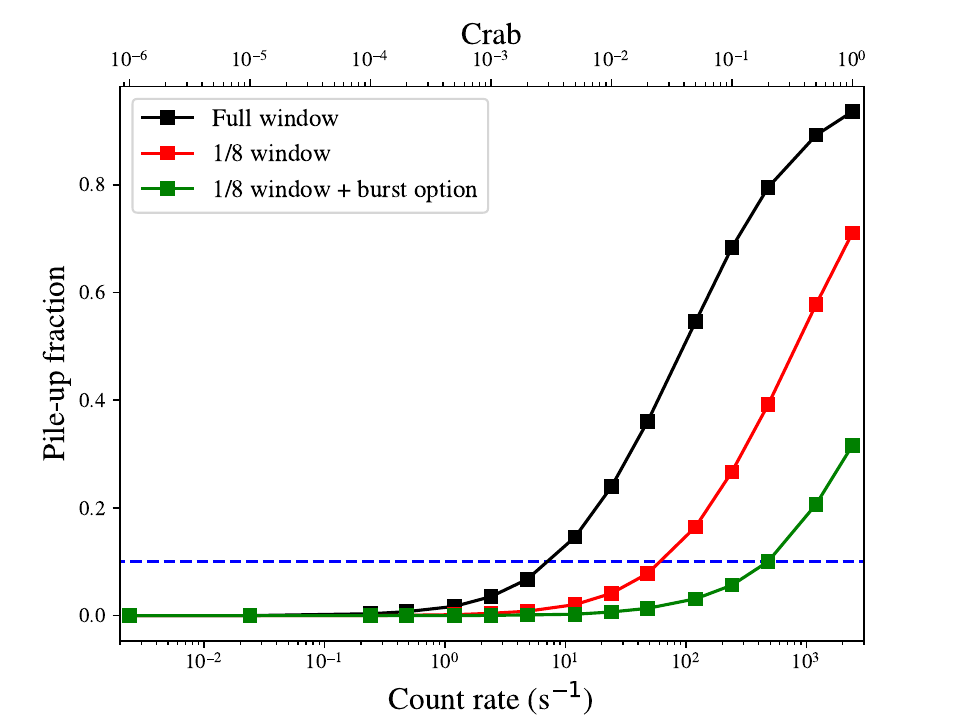}
   \end{center}
   \caption[Pile-up fraction for a point source with the Crab spectrum] 
   { \label{fig:point_crab} 
Pile-up fraction for each mode about intensity. Black, red, and Green plots correspond to full window mode, 1/8 window mode, and 1/8 window mode with 0.1\,sec burst option, respectively. The blue dashed line denotes the 10\% pile-up limit.}
\end{figure}

Assuming a point source, the model yields 2402.0\,counts\,s$^{-1}$ in the 0.4 -- 13.0\,keV energy range for Xtend without considering the pile-up effect. Hereafter, we define this as the ``1\,Crab'' spectrum and ``N\,Crab'' spectrum is then defined by scaling the normalization accordingly. For example, a $10^{-3}$\,Crab spectrum corresponds to a normalization of $8.9 \times 10^{-3}$\,counts\,s$^{-1}$, which yields 2.402\,counts\,s$^{-1}$ in the 0.4 -- 13\,keV band. We perform the pile-up simulation for an intensity range of $10^{-6}$ -- $1$\,Crab for each operation mode. The simulated spectra and the pile-up fractions are plotted in figure \ref{fig:point_crab_spec} and \ref{fig:point_crab}, respectively, from which the 10\% pile-up limit for each mode is derived (table \ref{tab:point_limit}). 
Here, the 10\% limit is set from the mission requirement. XRISM/Xtend is required to have an accuracy of 10\% for the effective area. Although the pile-up tolerance is not in the mission requirement, we set the same threshold for the effective area, because both the effective area and pile-up determine the accuracy for the flux measurement.

\subsubsection{Spectral hardening}
\begin{figure} [htbp]
   \begin{center}
   \begin{tabular}{c} 
   \includegraphics[width=10cm]{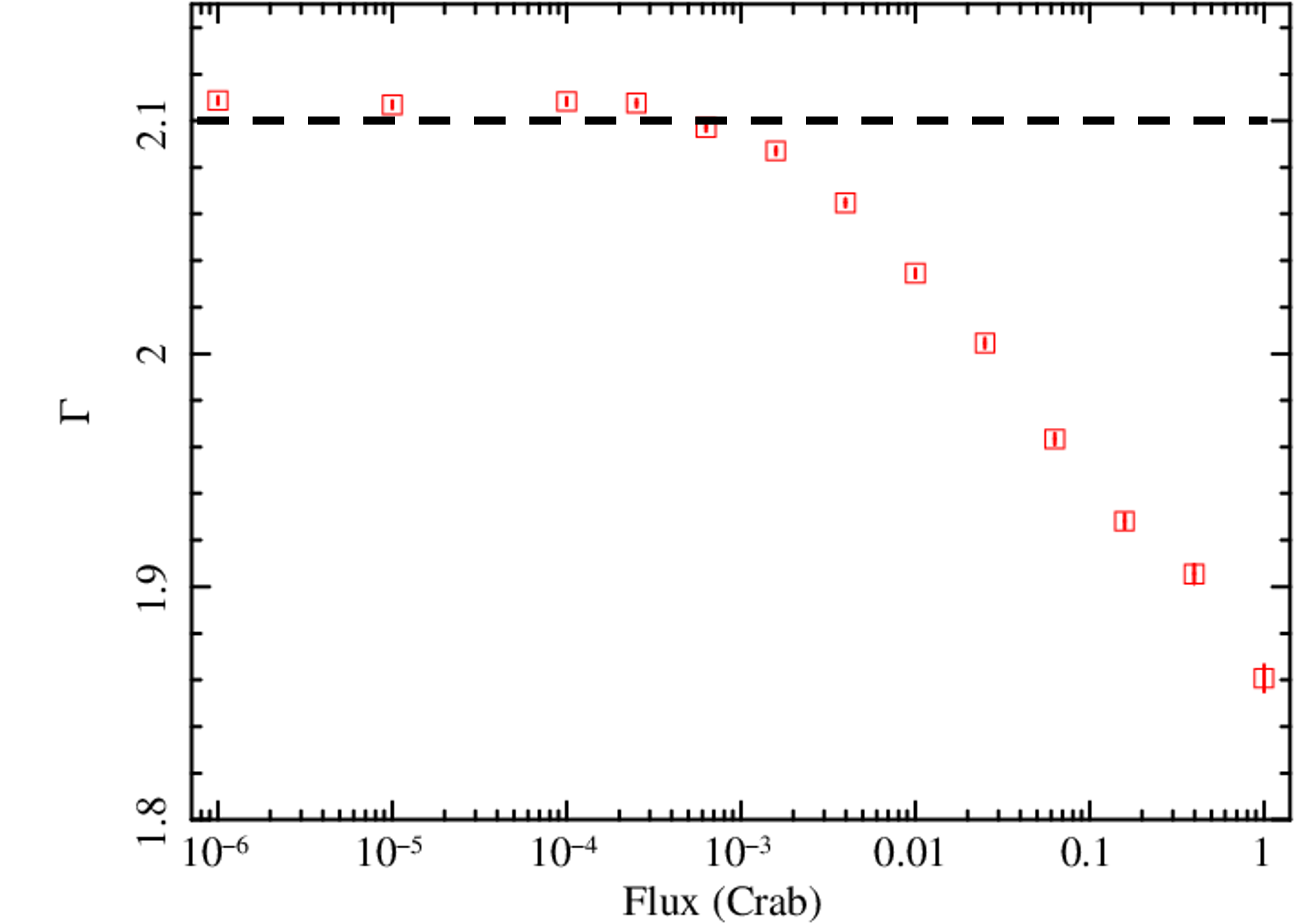}
   \end{tabular}
   \end{center}
   \caption[Spectral distortion of the piled-up spectra] 
   { \label{fig:crab_gamma} 
Photon indices ($\Gamma$) of the simulated spectra about the input flux for the full-window-mode observation of the Crab-like spectrum. The higher flux induces the heavier distortion of the spectral shape. The dashed line denotes the input photon index of 2.1.}
\end{figure}

When multiple photons emerge in a single pixel, they are detected as a single photon event with their merged energy, resulting in spectral hardening. From the simulated spectra, we estimated this effect. The full-window mode spectra are fitted with the input models themselves, whose photon indices and normalization are thawed. Figure \ref{fig:crab_gamma} shows the fitted photon indices along the input fluxes. The higher flux induces the heavier hardening of the spectral shape. At the 10\% pile-up limit, 3.2\,mCrab $=$ 7.8\,counts\,s$^{-1}$, the spectral hardening $\Delta\Gamma_{\rm pu}$ is $\sim$ 0.2, which is larger than the statistical error of $\Delta\Gamma_{\rm stat} \sim 0.06$.

\subsubsection{Grade migration}
\begin{figure} [htbp]
   \begin{center}
   \begin{tabular}{c} 
   \includegraphics[width=10cm]{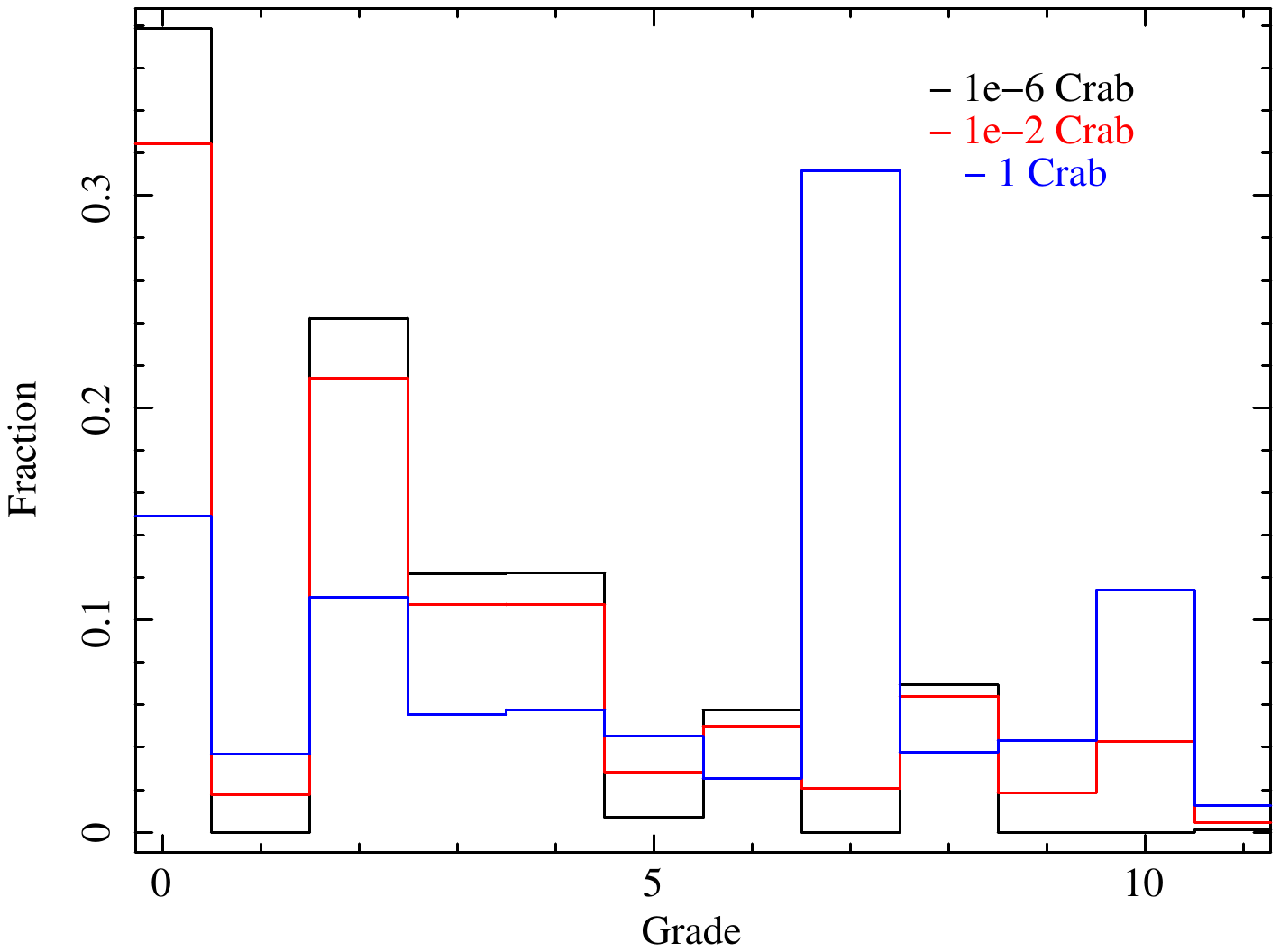}
   \end{tabular}
   \end{center}
   \caption[Grade branching ratios of the simulation] 
   { \label{fig:grades} 
Normalized Grade branching ratios for the simulated events. Black, red, and blue lines denote 1\,$\mu$Crab, 10\,mCrab, and 1\,Crab, respectively. Good grades are 0, 2, 3, 4, 6, and 8, from which X-ray spectra are extracted. Events with other grades are regarded as non-X-ray events.}
\end{figure}

When multiple photons emerge in adjacent pixels, they are detected as a bad-grade event with a non-X-ray pattern, which is the main cause of the flux underestimation. Figure \ref{fig:grades} shows the grade branching ratios of the simulated events. 1\,$\mu$Crab events, where pile-up does not matter, have a tiny fraction of bad grade events, whereas 1\,Crab events are dominated by bad grade of 7.

\subsubsection{Energy dependence}

The signal pattern of an X-ray event depends on its energy. Higher-energy photons produce larger signals; therefore, when their electron clouds are detected across multiple pixels, they are more likely to exceed the split threshold than lower-energy photons. When a hard X-ray spectrum is observed, it tends to have more multi-pixel events (grades $>$ 0) than those of a soft spectrum. Hence, the pile-up fraction depends on the spectral hardness, even if the count rate is the same. We then perform simulations for monochromatic spectra with energies of 1\,keV and 6\,keV. Here, ``1\,Crab'' spectrum in XSPEC is defined using a ``gauss'' model with a width of $\sigma = 0$\,eV and the normalization that yields 2402\,counts\,s$^{-1}$ for the Xtend's effective area. The ``N Crab'' spectrum is then defined analogously, following the same scaling as for the continuous spectra. The results are also summarized in table \ref{tab:point_limit}. Compared to the continuous Crab spectrum, the softer 1\,keV photons result in the slightly higher pile-up tolerance, and the harder 6\,keV photons the lower pile-up tolerance.

\begin{table}[htbp]
\caption{10\% pile-up limit for a point source} 
\label{tab:point_limit}
\begin{center}       
\begin{tabular}{cccc} 
\hline
 & The Crab & 1\,keV mono. & 6\,keV mono. \\
Operation mode & & Count rate (s$^{-1}$) / Intensity (mCrab) & \\
\hline
Full Window & 7.8 / 3.2 &  8.1 / 3.4 & 6.4 / 2.6 \\
1/8 Window & 66.2 / 27.5 & 69.7 / 29.0 & 55.0 / 22.9 \\
1/8 Window with 0.1\,s Burst & 447.9 / 198.9 & 504.5 / 210.1 & 395.2 / 164.5  \\
\hline
\end{tabular}
\end{center}
\end{table} 

\begin{figure} [htbp]
   \begin{center}
   \begin{tabular}{c} 
   \includegraphics[width=20cm]{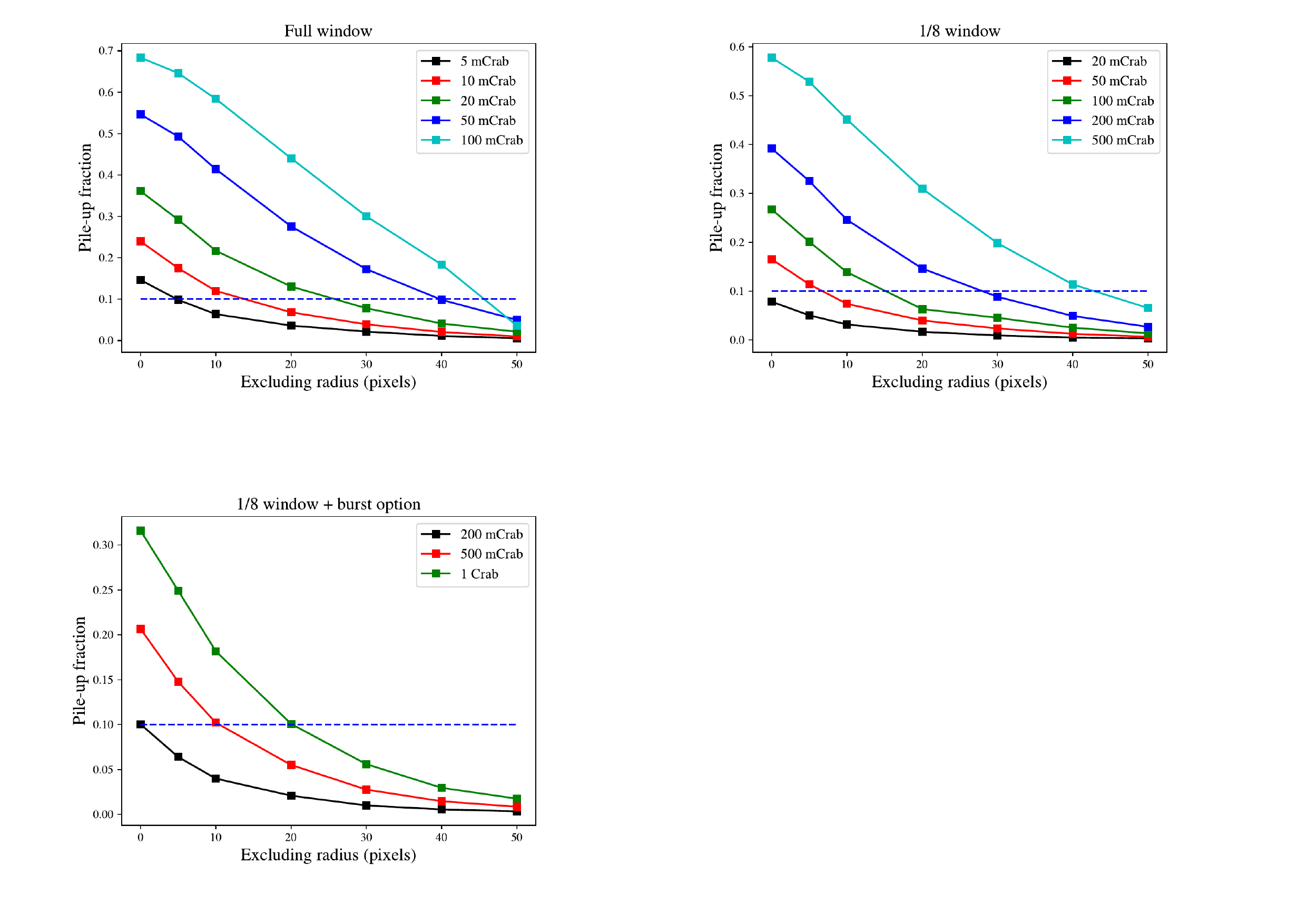} 
   \end{tabular}
   \end{center}
   \caption[Pile-up fraction for PSF core exclusion] 
   { \label{fig:exc_rad} 
Pile-up fraction for each mode about the PSF core exclusion radius in units of pixels. The top left, top right, and bottom left panels correspond to full window mode, 1/8 window mode, and 1/8 window mode with 0.1\,sec burst option, respectively. For each panel, the blue dashed lines denote the 10\% pile-up limit.}
\end{figure}

Pile-up occurs mainly in the PSF core, where most of the photons are accumulated. Hence, when an observation is significantly piled up, observers could exclude events around the PSF core to avoid pile-up, even if a significant fraction of events are rejected. We calculate the pile-up fraction about the excluded radius of the PSF for each mode. The results are plotted in figure \ref{fig:exc_rad}. One can see that, for example, when a bright source with the flux of 500\,mCrab is observed by the 1/8 window + burst mode, pixels near the PSF center are significantly affected by pile-up. Excluding the PSF core of 10\,pixels radius, the pile-up fraction is reduced to $\sim$ 10\%.

\subsection{Diffuse source}

\begin{figure} [htbp]
   \begin{center}
   \begin{tabular}{c} 
   \includegraphics[width=10cm]{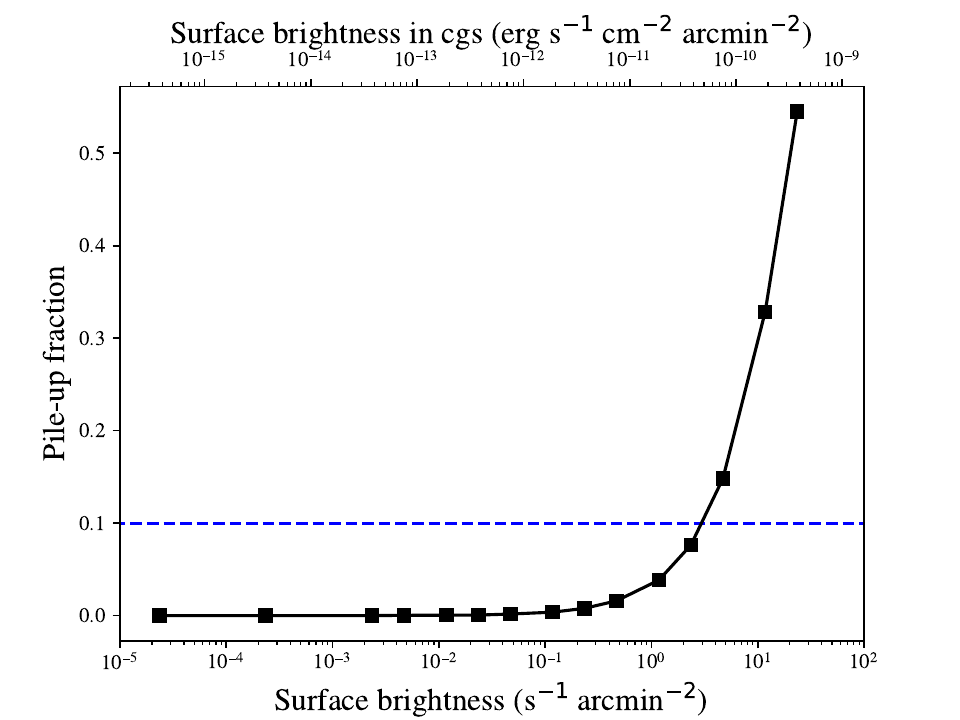}
   \end{tabular}
   \end{center}
   \caption[Pile-up fraction for a point source with the Crab spectrum] 
   { \label{fig:diffuse_crab} 
Pile-up fraction for the full window mode with regard to surface brightness. The blue dashed line denotes the 10\% pile-up limit.}
\end{figure}

Thanks to the large FOV and the low noise level, Xtend is suitable for diffuse sources. We perform simulations for a diffuse source with the Crab spectrum. A flat profile, i.e., a uniform surface brightness distribution, is assumed. Only the full window mode is considered, because the 1/8 window mode is not suitable for a diffuse source. The pile-up fraction about the surface brightness is plotted in figure \ref{fig:diffuse_crab}, from which the 10\% pile-up limit of 3.1\,counts\,s$^{-1}$\,arcmin$^{-2}$ (corresponds to $5.0 \times 10^{-11}$\,erg\,s$^{-1}$\,cm$^{-2}$\,arcmin$^{-2}$) is derived.

\section{Summary}

Pile-up is an unavoidable problem for photon-counting X-ray detectors when they observe a bright source. For {\it XRISM}/Xtend, we developed a pile-up simulator that calculates a pile-up fraction for an arbitrary spectrum and image of an X-ray source. The simulator is based on Monte Carlo simulations using the GEANT4 framework. We performed simulations for a point source with the Crab spectrum, monochromatic X-rays with energies of 1\,keV and 6\,keV, and a diffuse source with the Crab spectrum. The 10\% pile-up limit is estimated to be 7.8, 66.2, and 447.9 counts s$^{-1}$ for a point source with the Crab-like spectrum. Observers could refer to these results to select the Xtend operation mode when planning {\it XRISM} observations and to check whether their observations are piled up or not. Our results are public in the {\it XRISM} proposers' observatory guide\footnote{https://heasarc.gsfc.nasa.gov/docs/xrism/proposals/POG/}.

\section{Disclosures}
The authors declare that there are no financial interests, commercial affiliations, or other potential conflicts of interest that could have influenced the objectivity of this research or the writing of this paper.

\section{Code, Data, and Materials Availability}
The data that support the findings of this article are proprietary and are not publicly available. The data plotted in the above figures are available from the corresponding author upon request, and a limited subset of the underlying data can be requested from the corresponding author Tomokage Yoneyama.

\section{Acknowledgments}
We acknowledge the late Dr. Kiyoshi Hayshida for his significant contributions to the initial stage of these studies, especially for Tamba et al. (2022)\cite{tamba22}. This work is supported by the JSPS Core-to-core Program Grant Number JPJSCCA20220002, JSPS KAKENHI Grant Numbers 20K04009, 21H01095, 21K20372, 21K03615, 23H01211, 23K20850, 23K25907, and 24K00677. Major results of this paper was published as the SPIE proceedings, Volume 13093, id. 130935Y (2024).


\bibliography{report}   
\bibliographystyle{spiejour}   


\vspace{2ex}\noindent\textbf{Tomokage Yoneyama} is an assistant professor at Japan Aerospace Exploration Agency (JAXA). He received his BS, MS, and PhD degrees in science from Osaka University in 2016, 2018, and 2021, respectively.  His current research interests include neutron stars, X-ray instruments, and multimessenger time-domain astronomy.

\listoffigures
\listoftables

\end{spacing}
\end{document}